%% file: main.tex
\documentclass[10pt,twocolumn,letterpaper]{article}

\usepackage{cvpr}              % To produce the CAMERA-READY version
\input{preamble}
\definecolor{cvprblue}{rgb}{0.21,0.49,0.74}
\usepackage[pagebackref,breaklinks,colorlinks,allcolors=cvprblue]{hyperref}
\usepackage{algorithm, algpseudocode}
\usepackage{multirow}
\usepackage{graphicx} % for \rotatebox
\usepackage{layouts}
\usepackage[accsupp]{axessibility}

\def\paperID{41523} % *** Enter the Paper ID here
\def\confName{CVPR}
\def\confYear{2026}

\title{
Generative Diffusion Priors for 3D Mapping of the Dark Universe
}

\author{Brandon Zhao$^{1*}$\quad Diana Scognamiglio$^{2,3}$ \quad Olivier Dor\'e$^{2,4}$ \quad Katherine L. Bouman$^{1,4}$\\
\\
$^1$Department of Computing and Mathematical Sciences, California Institute of Technology\\  $^2$ 
Jet Propulsion Laboratory, California Institute of Technology\\ 
$^3$Department of Physics, Duke University\\
$^4$Cahill Center for Astronomy and Astrophysics, California Institute of Technology\\
{\tt\small $^*$byzhao@caltech.edu}
}

\begin{document}
\maketitle

\begin{abstract}
Reconstructing the three-dimensional distribution of dark matter from weak-lensing observations is a central but highly ill-posed inverse problem in cosmology. Unlike standard 3D reconstruction with multiple viewpoints, we observe the universe from a single line of sight, through noisy shape distortions of galaxies with uncertain distances, so meaningful recovery of the 3D matter field requires strong prior assumptions. Existing methods either produce point estimates with handcrafted priors or use neural ensembles for approximate Bayesian uncertainty, and struggle to capture the non-Gaussian, filamentary structure of the cosmic web. With the advent of new high-resolution cosmological simulations, we now have an alternative source of prior knowledge that captures the nonlinear statistics of structure formation with far greater fidelity than analytic prescriptions. We leverage these simulations to build a new dataset \texttt{Conicus3D}, which enables us to learn a data-driven diffusion-model prior capturing the full 3D distribution of dark matter structure across cosmic time. Building on recent plug-and-play approaches, we modify a diffusion-based posterior sampling scheme to the 3D weak-lensing setting, combining the learned prior with a differentiable physical forward model. On realistic simulations targeting a modern weak lensing survey, our approach yields substantially improved 2D and 3D reconstruction accuracy over baseline methods. Moreover, it produces posterior samples whose statistics closely track the underlying simulations, while remaining robust to moderate shifts in cosmology.

\end{abstract}

\section{Introduction}
Reconstructing the three-dimensional distribution of dark matter is a central challenge in modern cosmology. Because dark matter drives structure formation, recovering its 3D distribution is essential for testing gravity, constraining dark energy, and interpreting galaxy surveys to model the dynamics of the universe at large scales.

Dark matter does not emit or absorb light, so its structure must be inferred indirectly through its gravitational effects, most notably through the weak lensing (WL) of background galaxies. Recovering the underlying mass distribution from such indirect observations is an ill-posed inverse problem, and traditional reconstruction methods typically rely on smoothness-based priors which tend to overly damp high-frequency features. 

Recent advances in cosmological simulations now offer high-resolution, large-volume realizations of structure formation across cosmic time~\cite{perraudin2019cosmological, maksimova2021abacussummit, nelson2019illustristng, villaescusa2020quijote}. These datasets capture the full statistical complexity of the evolving dark matter field, enabling machine-learning approaches that learn directly from simulated universes. Modern generative models, in particular, provide powerful tools for representing these high-dimensional distributions as priors in inverse problems~\cite{zheng2025inversebench}.

Building on these developments, we introduce a new dataset, \texttt{Conicus3D}, and diffusion-based generative prior for 3D cosmological mass mapping. Our dataset, derived from state-of-the-art N-body simulations, consists of dark matter light cones paired with 2D lens-plane projections that preserve the relevant physics for lensing-based inference. Using this dataset, we train a score-based diffusion model that learns the 3D statistical distribution of cosmic structure and serves as a prior within a posterior sampler to generate physically realistic mass reconstructions.

\begin{figure*}
    \includegraphics[width=\textwidth]{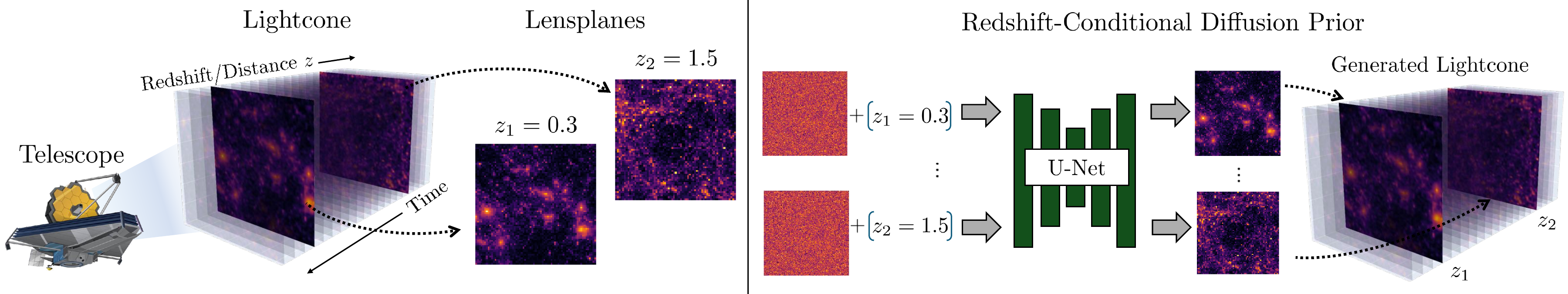}
    % \vspace{-.3in}
    \caption{Lightcone structure and redshift-conditional diffusion model. Left: The large-scale matter distribution is represented as a 3D lightcone, built from a sequence of lens planes slicing the dark matter density field at increasing comoving distance (redshift). Right: Each lens plane is modeled with a redshift-conditional 2D diffusion prior that learns the statistics of simulated overdensity maps at that redshift. During generation, we start from Gaussian noise on each plane, append one-hot redshift encodings, and denoise with a U-Net score model to obtain a coherent 3D lightcone. This redshift-conditioned factorization enables efficient generation of realistic volumetric dark matter fields consistent with lightcone geometry.}
    
    \label{fig:lightcone}
\end{figure*}

We demonstrate our method on the inverse problem of WL mass mapping, performing posterior sampling for tomographic reconstruction within simulations modeled after the James Webb Space Telescope (JWST) COSMOS-Web survey \cite{casey2023cosmos}. Our approach achieves substantially higher spatial fidelity than previous methods \cite{zhao2025revealing, simon2009unfolding}, produces posterior samples whose statistics match true simulations, and generalizes robustly to cosmologies beyond the training domain. Our key contributions are:
\begin{itemize}
\item \texttt{Conicus3D}: A first-of-its-kind 3D dark matter lightcone dataset, derived from state-of-the-art N-body simulations \cite{garrison2021abacus, maksimova2021abacussummit} and tailored for WL mass mapping, made publicly available for training and benchmarking by the computer vision and astrophysics communities;
\item A diffusion-based posterior sampling framework that combines data-driven priors with WL observables to enable robust, uncertainty-aware cosmological inference.
\end{itemize}

All data, simulated measurements, and code will be publicly released with documentation, establishing a reproducible testbed that invites broad participation in advancing 3D dark matter mass mapping.

\section{Background}
In the standard $\Lambda$CDM cosmological model, the universe began hot and nearly uniform, seeded by tiny quantum fluctuations in the primordial density field. As the universe expanded and cooled, gravity amplified those fluctuations into the cosmic web of dark matter halos, filaments, and voids that now host galaxies and gas. A central goal of cosmology is to determine how matter is arranged and how it evolves: the statistics of the initial conditions, the growth of structure over time, and the astrophysical processes that modulate it. Observationally, we infer this history indirectly, by measuring how light is deflected as it travels through the universe.

The focus of this work is a probe known as weak gravitational lensing, which measures the tiny, coherent ``shearing'' of galaxy shapes as their light passes through uneven distributions of matter. For a single galaxy, the induced distortion is subtle and masked by the galaxy's unknown intrinsic shape, but averaging over many galaxies recovers a shear field that statistically traces the underlying dark matter.

In cosmology, distance is often expressed in terms of redshift $z$, which measures how much cosmic expansion has stretched a galaxy’s light. Larger redshift corresponds to \textit{more distant} galaxies that we observe at \textit{earlier} cosmic times. Throughout this work, we treat redshift as a coordinate along the line of sight. Moreover, galaxy distances can be estimated from their colors via photometric redshifts; these redshift estimates provide a noisy but informative 3D signal along the line of sight that can be used to constrain the radial structure of the matter field.

Because light travels at a finite speed, increasing redshift also means looking deeper into the past. As a result, cosmological probes such as WL measure integrated signals along our past light cone, offering a layered view of the evolving matter distribution (see Fig. \ref{fig:lightcone}). 3D reconstruction therefore reveals not only where matter structures exist, but when they emerge in the history of the Universe.

\section{Related Work}
\paragraph{Weak Lensing Mass Mapping} 

Weak Lensing (WL) mass mapping is the inverse problem of reconstructing the distribution of total (mostly dark) matter along our past light cone from its gravitational-lensing imprints on galaxy images. In WL, deflections are small and can be approximated as stretching or magnification of intrinsically elliptical galaxy shapes. The induced shape change depends on a weighted line-of-sight projection of the matter overdensity, known as the convergence $\kappa$, which traces the integrated mass between the source galaxy and the observer. Because the cosmic matter field is dominated by dark matter, this task is referred to as \textit{dark matter mass mapping}.

Kaiser and Squires \cite{kaiser1993mapping} first showed that the 2D shear field $\gamma$, describing coherent shape distortions as a function of sky position, can be analytically inverted to yield a projected 2D convergence map of the intervening mass.
In practice, however, we only observe noisy estimates of $\gamma$  inferred from galaxy ellipticities, contaminated by each galaxy’s unknown intrinsic shape (``shape noise'').
Previous work proposes smoothing the observed galaxy shape field before inversion \cite{kaiser1993mapping}, while regularized approaches such as a Gaussian prior \cite{seljak1998weak} and sparsity \cite{lanusse2016high} have also been proposed. 
More recent developments include deep-learning-based reconstructions employing U-Net convolutional architectures \cite{jeffrey2020deep} and diffusion models \cite{remy2023probabilistic, boruah2025diffusion, boruah2025high}, which learn nonlinear priors directly from simulated data. Related simulation-based reconstruction work has also used diffusion models trained on simulation to probabilistically infer dark matter fields from biased galaxy tracers \cite{ono2024debiasing}.

While significant progress has been made in the development and implementation of 2D mass mapping algorithms, the problem of inferring the three-dimensional matter distribution from WL observations is extremely challenging and remains a largely unsolved problem. While observed galaxy redshifts (distances from the observer) can be estimated to give a 3D signal, the combination of intrinsic shape noise, large uncertainties in redshift estimation, and a single viewpoint limitation make the 3D mass mapping problem highly ill-posed. Existing approaches from the astrophysics community involve the use of strong regularizers such as a Gaussian prior \cite{simon2009unfolding}, sparsity constraints \cite{leonard2014glimpse}, singular vector truncation \cite{vanderplas2011three} or multiresolution wavelet methods \cite{massey2007dark} to obtain highly oversmoothed point estimates. 
Recent machine learning-based works \cite{zhao2024single, zhao2025revealing} have explored neural-field representations and ensemble models that provide approximate uncertainty quantification but do not correspond to exact Bayesian posterior sampling. Our approach targets the exact posterior with respect to the chosen simulation-derived prior.
Alternatively, rigorous Bayesian forward modeling approaches such as the BORG framework \cite{porqueres2021bayesian, porqueres2022lifting} infer the 3D density field by sampling initial conditions and evolving them through gravity solvers, ensuring physical consistency but at significant computational expense.
In this work, we introduce a diffusion-based framework that leverages outputs from state-of-the-art N-body simulations as physically motivated, data-driven priors. This approach produces high-resolution posterior samples of the 3D matter field consistent with both WL observations and structure-formation physics.
While we can use the mean of this posterior as a point estimate, being able to also sample the posterior will enable crucial uncertainty quantification as well as statistical estimation for downstream cosmological analyses. This direction is closely related to recent work using diffusion models trained on cosmological simulations for field-level emulation and parameter inference \cite{mudur2023cosmological, mudur2025diffusion}, but differs in that we infer the matter field indirectly through noisy WL measurements.

\paragraph{Diffusion Models for Inverse Problems}

Diffusion models have recently achieved state-of-the-art performance across a wide range of scientific inverse problems \cite{zheng2025inversebench}. A straightforward yet common approach for general scientific inverse problems treats a diffusion model as a posterior estimator, incorporating the measurements as conditioning through concatenation to the model input. However, this ignores the physics underlying the actual measurement model and thus can produce physically implausible solutions. 
From a Bayesian perspective, if the measurement process is defined by a known likelihood
$p(y | x)$ , the diffusion model acts as a score estimator for the data-driven prior
$p(x)$, approximating its log-gradient
$\nabla_x \log p(x)$.
To generate an unconditional prior sample $x_0$, a diffusion model smoothly removes noise from an unstructured initial noise sample $x_T$ by solving a stochastic differential equation. 
Many methods have been proposed for combining an unconditional diffusion model prior with a known measurement likelihood to sample from the resulting posterior. Early work such as DPS \cite{chung2022diffusion} uses guidance-based techniques to ``nudge'' samples along the likelihood gradient during training; however, the resulting samples are not theoretically guaranteed to correspond to a proper posterior. 
To address this limitation, a family of \textit{Plug-and-Play} \cite{venkatakrishnan2013plug} diffusion samplers has been developed, which incorporate the likelihood score term directly into the reverse process in a principled way. 
In this work, we propose a modified version of the DAPS (Decoupled Annealing Posterior Sampling, \cite{zheng2025inversebench}) framework, which unifies the learned diffusion prior with a known forward model to perform physically consistent posterior sampling. This enables us to reconstruct matter fields that fulfill observational constraints and preserve the statistics learned from cosmological simulations.

\section{\texttt{Conicus3D} - 3D Lightcone Dataset}

We introduce \texttt{Conicus3D}, a large-scale 3D lightcone dataset designed for training and evaluating dark matter mass-mapping algorithms on state-of-the-art, realistic N-body simulations of the universe. By providing a comprehensive and well-documented resource derived from high-fidelity cosmological simulations, our goal is to make the challenge of 3D WL mass reconstruction more accessible to both the astrophysics and computer-vision communities.
Upon release, \texttt{Conicus3D} will contain: 
\begin{itemize}
    \item 20,000 in-distribution dark matter lightcones, generated from a fiducial cosmology corresponding to the baseline parameters of the AbacusSummit suite \cite{garrison2021abacus, maksimova2021abacussummit}.
    \item 800 out-of-distribution lightcones, produced under alternative cosmological parameters that yield different statistical properties, enabling controlled evaluation of generalization performance across cosmological shifts.
    \item Mock galaxy shape catalogs, containing realistic galaxy positions, redshifts, and projection weights, along with open-source code for sampling additional shape catalogs to simulate observational noise and selection effects.
\end{itemize}

\paragraph{Dataset Construction} 

\texttt{Conicus3D} is derived from the AbacusSummit simulations \cite{garrison2021abacus, maksimova2021abacussummit}, which evolve over 300 billion particles under varying cosmological models. Each N-body simulation can be viewed as a function of (1) the cosmological parameters, which determine the statistical properties of the initial matter field and the cosmic expansion history, and (2) a random phase field, which sets the initial perturbations of the simulation. Evolving these initial conditions forward generates realistic projections of the dark matter field across redshift.

To create each lightcone volume, we process particle counts from the AbacusSummit base simulation, projecting the simulated particles directly onto the observer's lightcone geometry. We compute density contrasts (overdensities) for a sequence of redshift shells extending to $z \approx 2$, and bin the results into 20 lens planes equally spaced in comoving distance. Finally, we crop each plane to the COSMOS survey footprint so that the resulting lightcone volumes match the geometry and sky coverage of real observations.

\paragraph{Broader Applications}
While our dataset is primarily designed for 3D dark matter mass mapping, its structure makes it naturally extensible to a wide range of inverse problems on the lightcone. Many astrophysical observables---such as Fast Radio Bursts (FRBs) \cite{petroff2019fast}, CMB secondary anisotropies \cite{aghanim2008secondary, carlstrom2002cosmology}, and 21-cm intensity maps \cite{bull2015late}---can be interpreted as line-of-sight projections of the underlying matter field across cosmic time. The same lightcone geometry, redshift binning, and comoving-distance sampling used in this work provide a unified framework for simulating and reconstructing these observables. 

\begin{figure}
    \includegraphics[width=0.48\textwidth]{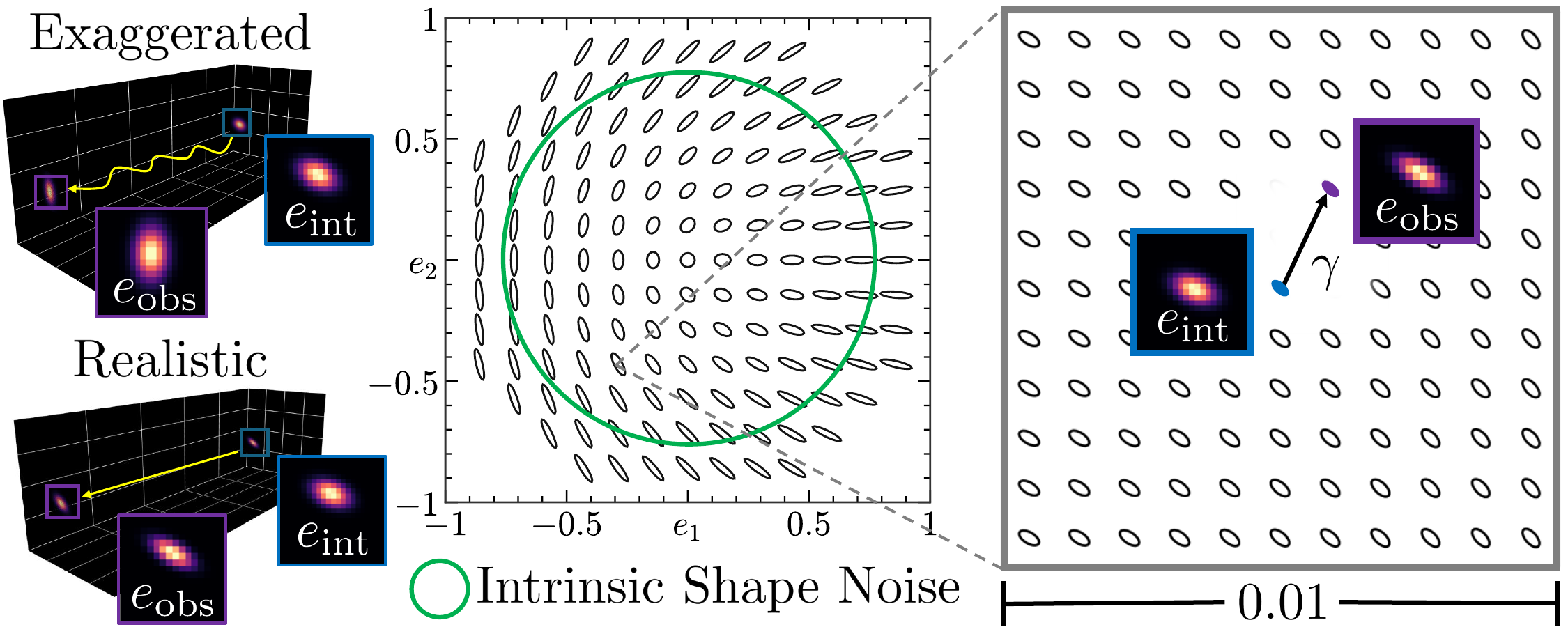}
    \caption{Weak Lensing Measurements. (Left) As light from distant galaxies propagates through the universe, it is deflected by intervening matter, producing small differences between the intrinsic (unlensed) shape $e_\text{int}$ and the observed shape $e_\text{obs}$. These subtle distortions provide a powerful probe of the 3D dark matter distribution. (Right) To characterize cosmic shear, galaxy shapes are often represented in the complex ellipticity plane, whose components encode the axis ratio and orientation of an ellipse. In typical WL surveys, shape noise is more than two orders of magnitude larger than the shear signal, so the lensing effect, described by the shear $\gamma$ and approximately additive in ellipticity space, must be extracted statistically by combining many measurements from a dense field of galaxies to recover the coherent imprint of the underlying matter distribution. Figure adapted from \cite{zhao2025revealing, schneider2006weak}.}
    \label{fig:shape_change}
\end{figure}

\section{Methods}

\subsection{Weak Lensing Measurements}
As light from distant galaxies travels toward us, its trajectory is continuously deflected by the intervening (dark-dominated) matter distribution.
In the WL regime, these deflections are small and manifest as slight distortions in the observed galaxy shapes.
To quantify these small changes, approximately elliptical galaxy shapes are described using a complex ellipticity parameter $e$. Given an approximately elliptical galaxy with axis ratio $r$ and orientation angle $\phi$, we can define the magnitude and phase of its ellipticity $e$ as in \cite{schneider2006weak}: 
\begin{equation}
    \label{eq:ellipticity}
  |e| = \frac{1-r}{1+r} \; \hspace{0.5in} \angle e = 2\phi.
\end{equation}
WL measurements relate a galaxy's observed ellipticity $e_\text{obs}$ to its intrinsic ellipticity $e_\text{int}$, which describes what its observed shape would be in the absence of lensing. In the regime of WL, the lensing effects of intervening matter can be approximately described as a shear $\gamma$, which is additive in the ellipticity domain (\cite{schneider2006weak}): 
\begin{equation}
\label{eq:shapechange}
  e_\text{obs} - e_\text{int} = \gamma.
\end{equation}
An overview of shear measurements and the complex ellipticity can be found in Fig. \ref{fig:shape_change}. In practice, although the observed shape $e_\text{obs}$ can be measured directly from an image, a galaxy's intrinsic shape $e_\text{int}$ is much more difficult to estimate. In traditional WL surveys, $e_\text{int}$ is completely unknown \cite{schneider2006weak}. % and represented as an additive Gaussian ``shape noise''.  
The standard assumption is that the true shape of galaxies are randomly distributed without a preferred overall direction, so $e_\text{int}$ is modeled as an additive Gaussian ``shape noise''.
To mitigate the effects of the shape noise, WL surveys combine measurements from many galaxies into a galaxy catalog containing the sky position, redshift (distance), and observed shapes of millions of galaxies, constituting the measurements for 3D mass mapping.

The shear $\gamma$ observed in a survey image arises from the cumulative lensing occurring along the line of
sight between us and the galaxy. More precisely, the shear is a projection followed by a 2D convolution of the {\it overdensity} $\delta = (\rho - \bar{\rho} ) / \bar{\rho}$ intersected by a ray of light as it travels towards earth. Here $\rho$ is the continuous matter density field and $\bar{\rho}$ is the mean density of the universe at the time of the intersection. In this section we introduce only the specific equations governing the forward model; a more detailed derivation can be found in the supplement.
% \vspace{-.22in}
\paragraph{Projection} To compute the shear from the matter overdensity field $\delta$, we first introduce the convergence $\kappa$ \cite{schneider2006weak}, which for a source galaxy with sky position $\theta$ and comoving distance\footnote{Comoving distance refers to a distance measure that factors out the expansion of the universe.} $w$ is: 
\begin{equation}
    \label{eq:proj}
    \kappa(\boldsymbol{\theta}, w) = \mathbf{Q}\delta = \frac{3H_0^2 \Omega_m}{2c^2} \int_0^w \mathrm{d}w' \frac{w'(w-w')}{w} \frac{\delta(w'\boldsymbol{\theta}, w')}{a(w')},
\end{equation}
where $H_0$ and $\Omega_m$ are cosmological parameters and $c$ is the speed of light. The scale parameter $a(w)$ represents the known expansion of the universe that occurs as light travels towards us, and $\delta$ is the matter overdensity. In the case of standard WL surveys, $w$ has a degree of uncertainty, as it is inferred from photometric estimates of the galaxy redshift. In this case, we take the convergence to be the expected value over the estimated distribution of the galaxy comoving distance $w$, thereby incorporating photometric redshift uncertainty directly into the forward model.

\paragraph{Convolution} $\gamma$  is obtained by convolving the convergence $\kappa$ with a complex kernel $\mathcal{D}(\boldsymbol{\theta}) = -1/(\boldsymbol{\theta^*})^2$ (\cite{schneider2006weak}): 
\begin{equation}
    \label{eq:conv}
    \gamma(\boldsymbol{\theta}, w) = \mathbf{P}\kappa = \frac{1}{\pi} \int_{\mathbb{C}^2} \mathrm{d}^2 \boldsymbol{\theta'} \mathcal{D}(\boldsymbol{\theta} - \boldsymbol{\theta'})\kappa(\boldsymbol{\theta'}, w).
\end{equation}
Here, the angular position $\boldsymbol{\theta} = \theta_1 + i\theta_2$ is expressed in complex coordinates, and the asterisk $*$ represents complex conjugation. Notably, the forward operators in Eqns. \ref{eq:proj} and \ref{eq:conv} are both linear with respect to the overdensity $\delta$, so obtaining the overdensity from the shear measurements is a linear inverse problem. Combining Eqns.~\ref{eq:shapechange}, \ref{eq:proj} and \ref{eq:conv} gives the full formulation with the effects of the additive shape noise $\varepsilon$:
\begin{equation}
    \label{eq:forward_model}
    e_\text{obs} = \mathbf{P} \mathbf{Q}\delta + \varepsilon \; \hspace{0.5in} \varepsilon \sim \mathcal{N}(0, \sigma_{\text{shape}}).
\end{equation}
This assumption of additive Gaussian noise implies a Gaussian likelihood for the observed ellipticities across the galaxy catalog.
Reconstructing $\kappa = \mathbf{Q}\delta$ rather than the full 3D overdensity is known as 2D mass mapping, and is much more well-constrained. A majority of the uncertainty in 3D mass mapping comes from the inversion of the projection operator $\mathbf{Q}$. Although some resolution in the radial direction can be inferred using the known distances to (i.e., redshifts of) galaxies, uncertainty in these redshift estimations coupled with shape noise and single-view geometry make the backprojection severely ill-posed.

\begin{figure*}
    \centering
    \includegraphics[width=\textwidth]{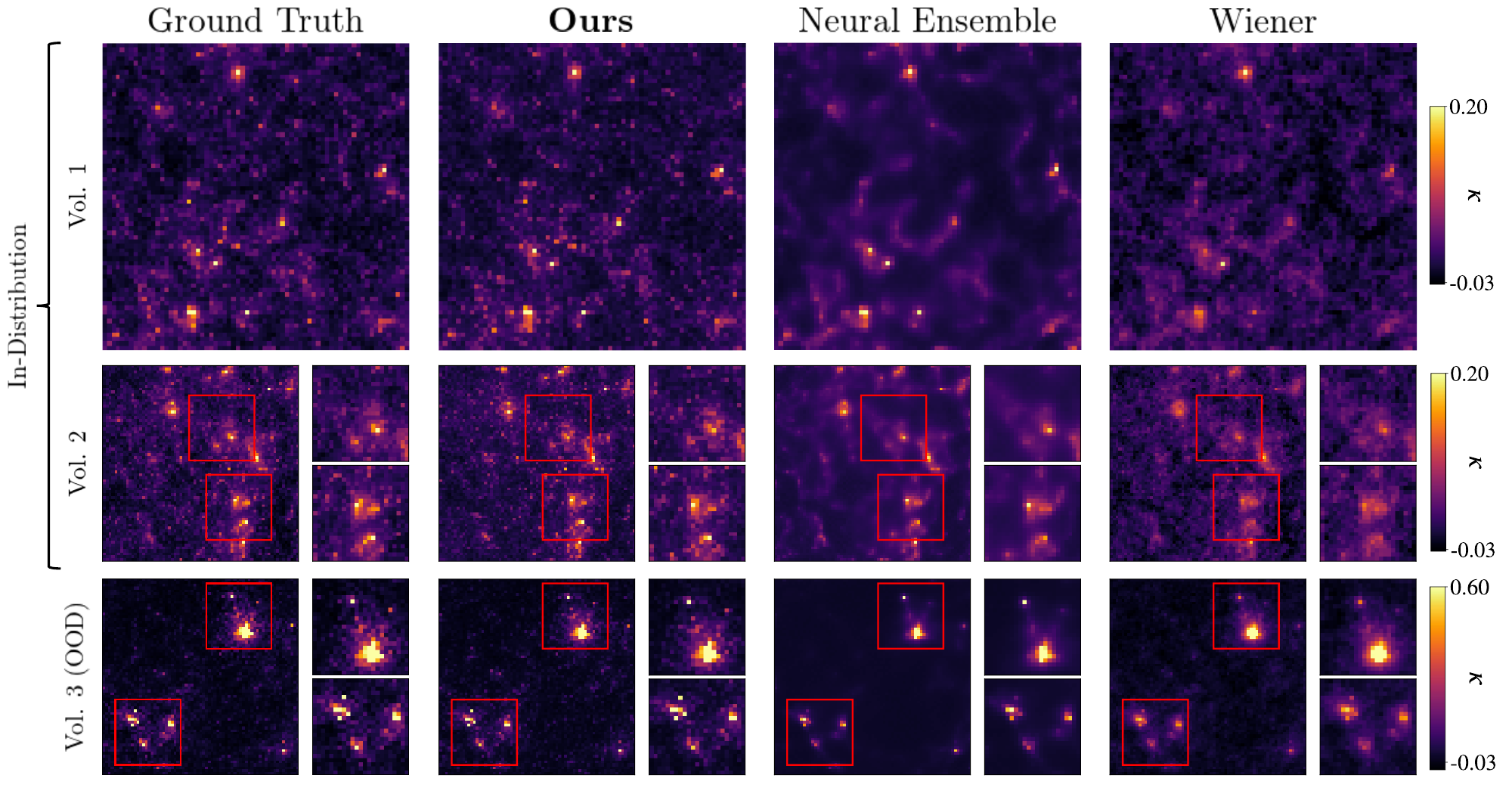}
    % \vspace{-.4in}
    \caption{Two-dimensional mass reconstruction results from simulated WL data. For three representative lightcone volumes (rows), we show the projected convergence ($\kappa$) maps for the ground truth, our method, the Neural Ensemble baseline, and the Wiener filter reconstruction (columns); for each method, we use the posterior mean as a point estimate. Our reconstructions recover sharp cluster peaks and filamentary features that closely match the true $\kappa$ maps, while the baselines exhibit oversmoothing and reduced contrast in high-density regions. The third row highlights performance under a cosmology mismatch; even in this out-of-distribution setting, our method maintains high-fidelity 2D mass reconstructions that better preserve the locations and amplitudes of prominent structures. }
    \label{fig:recon2d}
    % \vspace{-.15in}
\end{figure*}

\begin{figure*}
    \centering
    \includegraphics[width=\textwidth]{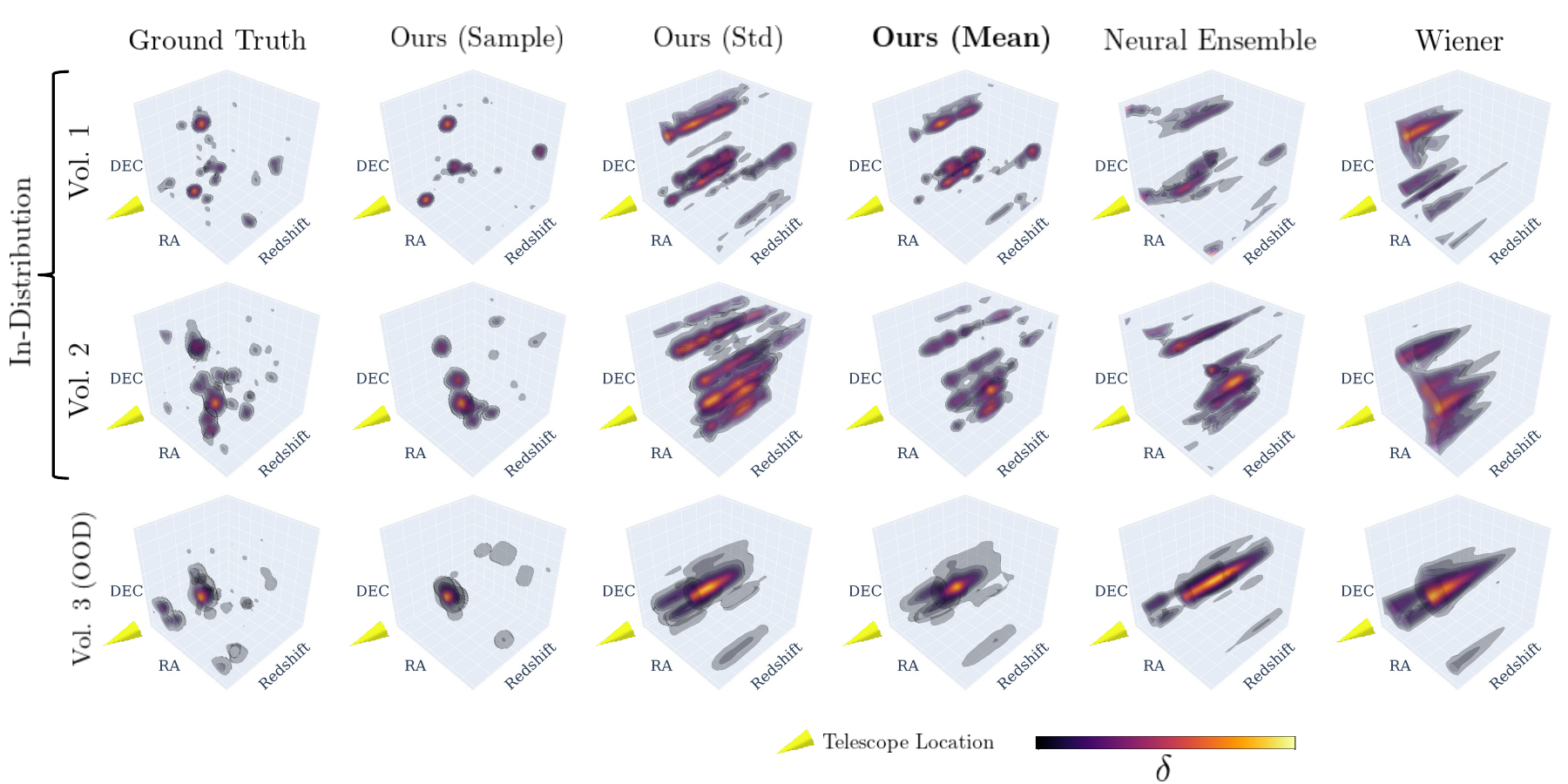}
    % \vspace{-.4in}
    \caption{Three-dimensional mass reconstruction results from simulated WL data. Each row corresponds to a different simulated lightcone volume, and columns compare the ground-truth overdensity field to reconstructions from our diffusion-based posterior sampler, the Neural Ensemble baseline \cite{zhao2025revealing}, and the Wiener filter method \cite{simon2009unfolding}; for each method, we use the posterior mean as a point estimate. Our reconstructions better recover the spatial morphology of clusters and filaments and place structures at the correct redshift slices, whereas the baselines tend to smear overdensities along the line of sight or miss small-scale features. The third row illustrates the challenging case of a misspecified cosmology with massless neutrinos, where our method still tracks the true 3D structure more faithfully despite a mismatch between the prior cosmology and the ground-truth simulation.}
    \label{fig:recon3d}
    
\end{figure*}

\subsection{Lightcone Diffusion Model}

Diffusion models are powerful generative models which generate data by learning to reverse a diffusion noising process applied to the data distribution. In this work, we derive our data-driven prior by estimating the score of the data distribution $p(\delta)$ of possible lightcones $\delta$. As is common in WL analyses, we represent each lightcone $\delta$ as a set of discrete lensplanes $\{\delta^{(z)}\}_{z=1}^M$, and take the distributions of lensplanes at different redshifts to be independent, i.e. $p(\delta) = \prod_{z=1}^Mp(\delta^{(z)})$. Because the lensplanes cover such large distance scales, this is a reasonable assumption to make \cite{simon2009unfolding}; this can also be seen empirically in the simulated dataset (see Fig. \ref{fig:ps_kappa}).
We then train a conditional diffusion model using \texttt{Conicus3D} to estimate the prior distributions $p(\delta^{(z)} | z)$ for each of $M$ redshift values, which are fixed beforehand. Our diffusion model uses a standard DDPM UNet architecture operating on $128 \times 128$ lens-plane images. In this way, we can use a traditional image denoising model to generate 3D lightcones, by denoising a batch of images with a list of one-hot vectors corresponding to different redshifts; see Fig. \ref{fig:lightcone} for an illustration.

\subsection{Decoupled Annealing Posterior Sampling}

Given our data-driven lightcone prior $p(\delta)$, and the physics-based likelihood $p(\gamma|\delta)$ from the forward model above, we wish to sample from the resulting posterior distribution
\begin{equation}
    p(\delta | \gamma) \propto p(\gamma|\delta)p(\delta).
\end{equation}
To do so, we propose a modified version of the Decoupled Annealing Posterior Sampling (DAPS) algorithm, which has been shown to be successful on a wide range of scientific inverse problems in the InverseBench framework \cite{zhang2025improving}. DAPS samples a target ``clean'' posterior $p(\delta_0 | \gamma)$ by sequentially generating samples from noise annealed posteriors $p(\delta_t | \gamma)$ from $t = T$ to $0$. To generate samples $\delta_t$ from $\delta_{t+\Delta t}$, DAPS alternates between (1) sampling $\delta_{0|\gamma} \sim p(\delta_0 | \delta_{t+\Delta t}, \gamma)$, and (2) sampling $\delta_t \sim \mathcal{N}(\delta_{0 | \gamma}, \sigma^2_{t_2}\mathbf{I})$. This process is repeated for decreasing $t$ until $\delta_0$ are sampled from the target posterior \cite{zhang2025improving}.

The original DAPS algorithm samples $p(\delta_0 | \delta_t, \gamma) \propto p(\delta_0 | \delta_t) p(\gamma | \delta_0)$ with score-based methods such as Langevin dynamics, approximating the intractable $p(\delta_0 | \delta_t)$ with a Gaussian distribution with diagonal covariance $\mathbf{\Sigma}_t$. While this approximation enables efficient posterior sampling, it does not encode any information about possible spatial correlations that may exist in $\delta$. Motivated by the cosmological principles of translation invariance and isotropy, we propose a physically informed covariance which is instead diagonalized in Fourier Space according to the matter power spectrum (PS) $P_k$: 
% \vspace{-.1in}
% \begin{equation}
%     \mathbf{\Sigma} = F^{-1}\text{diag}(P_k)F.
% \end{equation}

\begin{equation}
        \mathbf{\Sigma}_t^{-1}
        =
        F^{-1}\operatorname{diag}\left(\sigma_t^{-2} + P_k^{-1}\right)F .
\end{equation}

In our experiments, we estimate $P_k$ empirically by averaging over the training data. 
This Fourier-diagonal covariance structure arises in any inverse problem over approximately stationary fields, including adaptive optics~\cite{lane1992simulation}, geophysical inversion~\cite{hansen2006linear}, and CMB map-making~\cite{elsner2013efficient}, suggesting that our spectral sampling strategy could improve diffusion-based posterior sampling in these settings; see Supplement Sec.~\ref{sec:spectral_applicability} for further discussion.
A detailed explanation and pseudocode of our specific implementation can also be found in the supplement.

While the posterior mean provides a strong map-level reconstruction, most WL science evaluates non-linear functionals of the field posterior. Our sampler targets the full posterior, combining a data-driven diffusion prior with the physical likelihood so that each draw preserves small-scale power and realistic radial structure. This ensures that statistics computed from the ensemble---not only from the mean---are unbiased and well calibrated, which is crucial when moving from simulations to real data.

\section{Experiments}

\begin{figure*}
    % \vspace{-.45in}
    \includegraphics[width=\textwidth]{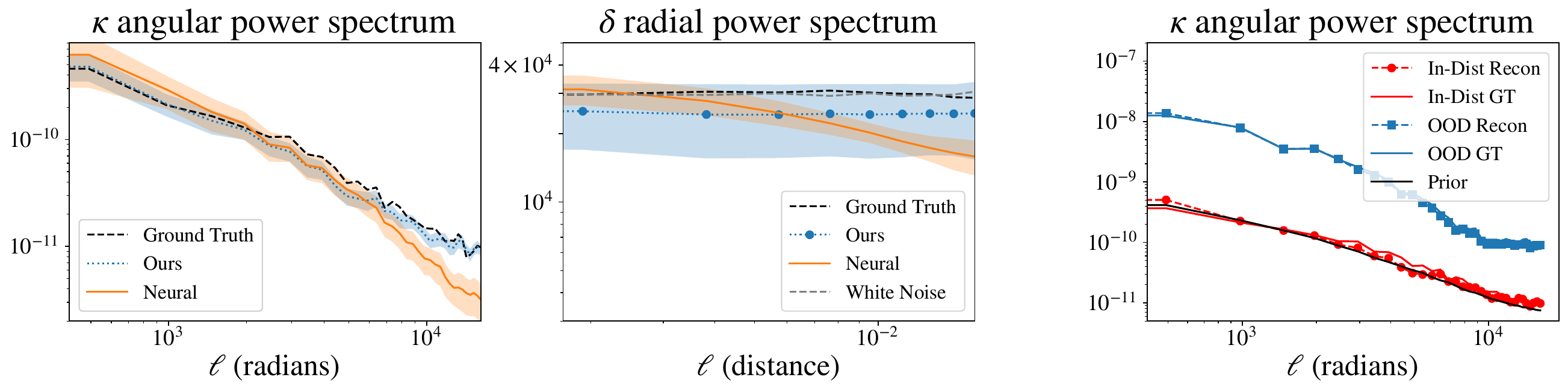}
    \caption{Sample quality. Our diffusion-based posterior sampler produces 3D mass-field realizations whose statistics match those of realistic dark matter simulations. Left: Angular PS $C_\ell^\kappa$ of the convergence $\kappa$ for individual samples. Our samples retain the correct high-$\ell$ (small-scale) power, whereas the Neural Ensemble \cite{zhao2025revealing} baseline over-smooths small-scale modes when the noise power exceeds the signal power and the reconstruction becomes prior-dominated. Center: Radial PS of the overdensity $\delta$ across lens planes. Our samples are nearly flat in this radial spectrum, indicating decorrelation between different redshift planes and avoiding the spurious line-of-sight correlations seen in the neural ensemble outputs. Because many downstream WL statistics are nonlinear functionals of each sample (e.g., peak counts, voids), matching these sample-level power spectra is essential for unbiased cosmological inference and calibrated uncertainties. Right: Our diffusion prior is trained under a fiducial cosmology, which implies an expected angular PS for $\kappa$ (``Prior''). We plot five power spectra: the ground truth and reconstruction for an in-distribution cosmology (In-Dist GT / In-Dist Recon), the ground truth and reconstruction for an out-of-distribution cosmology with massless neutrinos (OOD GT / OOD Recon), and the prior from our diffusion model. When the true cosmology matches the training cosmology, prior samples and reconstructions reproduce the same spectrum. In the out-of-distribution case, the true power is significantly higher, but our posterior samples still match the ground-truth spectrum due to likelihood guidance, showing that the method adapts to the observed universe even when the prior cosmology is misspecified.}
    
    \label{fig:ps_kappa}
    \label{fig:ood_ps}
\end{figure*}
\subsection{Mock Weak Lensing Survey}
\label{sec:recon}

We evaluate our method on simulated WL data modeled after the JWST COSMOS-Web survey \cite{casey2023cosmos}, which targets one of the most extensively studied extragalactic fields, large enough to capture a representative portion of the cosmic large-scale structure. While the shear catalogs for JWST COSMOS-Web are not publicly available, we consider mock surveys that reproduce the footprint, source density, and shape-noise characteristics of the real COSMOS-Web observations \cite{scognamiglio2025highest}. We assume a footprint of 0.54 deg$^2$ with a mean redshift catalog source density of $\approx$ 261 galaxies arcmin$^{-2}$, an intrinsic shape-noise dispersion of $\sigma_e \approx 0.25$ \cite{scognamiglio2025highest}, and Gaussian photo-$z$ dispersion $\sigma_z = 0.11(1 + z)$.
We compare against two publicly available baseline methods for 3D mass mapping: the analytical Wiener filter reconstruction \cite{simon2009unfolding} and the neural ensemble estimator \cite{zhao2025revealing}. 

We first present results for recovering the projected 2D convergence field $\kappa$ (a weighted integral of the overdensity $\delta$ along the line of sight) from a simulated light cone. Figure \ref{fig:recon2d} and Table \ref{tab:recon} show the reconstructed $\kappa$ maps and quantitative comparisons. For each method, we compute the posterior mean projected overdensity across 128 samples and evaluate agreement with the ground-truth map using the mean-subtracted Pearson cross-correlation coefficient, which is appropriate since the WL forward model is invariant under constant additive shifts in $\kappa$.
Our method achieves significantly higher correlation with the true convergence field than both the Wiener filter and neural ensemble baselines, demonstrating the ability of the diffusion-based posterior sampling framework to recover fine-scale structure while maintaining physically consistent large-scale modes.
% \vspace{-.05in}
In Fig. \ref{fig:recon3d} we reconstruct the 3D overdensity field $\delta$ for the same simulated light cones up to redshift $z=1$. We again compute the posterior mean across 128 samples as a point estimate of the overdensity. Because of single-view limitations, and large amounts of noise, it is only possible to reconstruct the overdensity field with a limited radial resolution. Uncertainty in the radial position of structures results in point estimates with a blur along the line of sight. For each method, we additionally blur the ground truth and reconstructed volumes with a Gaussian filter of length $\sigma = 4$ lensplanes in the radial $z$ direction (see supplement for detailed analysis of $z$ blur lengths). We then report the cross-correlation (averaged across lensplanes) of each reconstruction with the ground truth, blurred and unblurred. Our method outperforms the 3D reconstruction baselines in each volume. 
\begin{table}
    \centering
    \small
    \begin{tabular*}{0.48\textwidth}{@{\extracolsep{\fill}}ccccc}
        \toprule
        Vol.&Method & $\rho_\text{blur}^{\text{3D}}$($\uparrow$) & $\rho^{\text{3D}}$($\uparrow$) & $\rho^\text{2D}$($\uparrow$) \\
        \midrule
        \multirow{3}{*}{1} 
        & \textbf{Ours}  & \textbf{0.83} & \textbf{0.23} & \textbf{0.87} \\
        & Neur. \cite{zhao2025revealing} & 0.79 & 0.21 & 0.86 \\
        & Wien. \cite{simon2009unfolding}   & 0.71 & 0.21  & 0.77 \\
        \midrule
        \multirow{3}{*}{2}
        & \textbf{Ours}    & \textbf{0.83} & \textbf{0.27} & \textbf{0.88} \\
        & Neur. \cite{zhao2025revealing} & 0.80 & 0.21 & 0.87 \\
        & Wien. \cite{simon2009unfolding}   & 0.72  & 0.23  & 0.83 \\
        \midrule
 \multirow{3}{*}{3}& \textbf{Ours} & \textbf{0.92} & \textbf{0.18} & \textbf{0.98}\\
         & Neur. \cite{zhao2025revealing} & 0.86 & 0.09 & 0.96 \\
         & Wien. \cite{simon2009unfolding} & 0.84 & 0.13 & 0.92\\
         \bottomrule
    \end{tabular*}
    % \vspace{-.1in}
    \caption{Quantitative comparison of reconstruction quality. We summarize 3D and 2D mass-mapping performance for three simulated JWST-scale lightcone volumes.  We report the Pearson cross-correlation coefficients averaged over lensplanes between reconstructed and ground-truth maps: $\rho_\text{blur}^{\text{3D}}$ where both lightcones are first blurred with a Gaussian kernel of width $\sigma = 4$ lensplanes in the line-of-sight direction, $\rho^{\text{3D}}$ for the full-resolution 3D fields, and $\rho^\text{2D}$ for the projected 2D convergence maps (higher is better in all cases). Our diffusion-based method consistently achieves higher correlations than both the Neural Ensemble and Wiener filter baselines across all three volumes in 2D and 3D.}
    \label{tab:recon}
\end{table}

\subsection{Cosmology Mismatch}
Because our method is supervised, a key question is whether or not we can generalize well under a reasonable error in our prior assumptions. Thus, we repeat the experiment described in Sec. \ref{sec:recon} but for a ground truth lightcone generated under different cosmological assumptions. In particular, we consider a cosmology with massless neutrinos, which results in significantly higher amplitudes in the PS. Recovery results are shown in row 3 of Figs. \ref{fig:recon2d}, \ref{fig:recon3d}. Our method generalizes well despite a mismatch in the statistics of the prior and ground truth volumes. In this case, the OOD cosmology produces higher power and thus stronger lensing signal, allowing the likelihood to partially overcome the mismatched prior.

In addition, we compare the PS of both the fiducial (assumed) cosmology and the out-of-distribution cosmology, the PS for both of our reconstructions, and the PS for unconditional PS in Fig. \ref{fig:ood_ps}. Although our prior samples match the fiducial cosmology of the training set, our method is able to generate posterior samples with angular PS matching the ground truth due to likelihood guidance.

\subsection{Sample Quality}
Beyond point estimation, our method's principal advantage is the ability to \textit{sample} from a realistic posterior whose statistics are consistent with the true dark matter distribution.
We assess sample realism using the angular PS of the recovered 2D convergence maps (Fig. \ref{fig:ps_kappa}); matching both amplitude and scale-dependence indicates correct two-point statistics.

Although all methods have access to the same theoretical PS, smoothness regularization in the Neural Ensemble estimator inevitably damps small-scale structure in each reconstruction. This suppression wipes out high-frequency power and produces convergence maps that look overly smoothed (Fig. \ref{fig:ps_kappa}). In contrast, our diffusion-based posterior generates high-resolution samples whose angular power spectra match the ground truth across all scales.
This fidelity is essential for extracting reliable cosmological information from WL data. Many key probes, such as peak and void counts \cite{kratochvil2010probing}, Minkowski functionals \cite{kratochvil2012probing}, bispectra \cite{cooray2001weak}, halo and filament finders \cite{knebe2011haloes, sousbie2011persistent}, cross-correlation with galaxies, clusters \cite{chang2015wide, vikram2015wide}, or CMB lensing \cite{hand2015first}, depend directly on the detailed small-scale and radial structure in each possible 3D realization of the matter field. If the samples are oversmoothed or exhibit spurious correlations across redshift planes, the resulting cross-correlation amplitudes, cosmological parameter constraints, and covariance estimates will be systematically biased.
Our samples robustly reproduce the correct angular power and remain decorrelated along the line of sight, ensuring that downstream non-Gaussian statistics remain unbiased and that ensemble-based uncertainty propagation is physically meaningful.

\begin{figure}
    \includegraphics[width=0.45\textwidth]{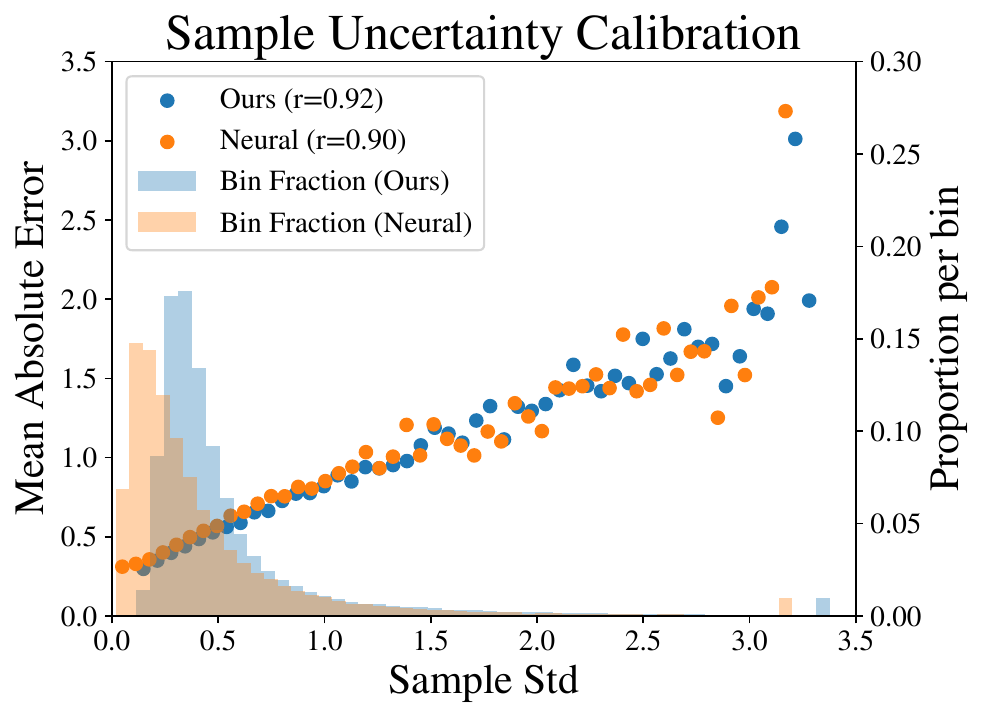}
    \caption{Uncertainty calibration. For each voxel, we compute the standard deviation of the posterior samples (“sample std”) and the mean absolute error (MAE) between the posterior mean and the ground truth. We then bin voxels by sample std and plot the average MAE per bin, along with the fraction of voxels in each bin. Both our method (r = 0.92) and the neural ensemble estimator (r = 0.90) \cite{zhao2025revealing} show a strong correlation between predicted uncertainty and actual error, but only our samples correspond to a well-defined Bayesian posterior distribution.}

    \label{fig:calibration}
\end{figure}

\subsection{Uncertainty Calibration}

Another important aspect of our method is the ability to estimate uncertainty from the statistics of samples recovered from our posterior sampler. Results are shown in Fig. \ref{fig:calibration}. We show a very clear correlation between sample standard deviation and error with the ground truth, with a near perfect correspondence for a majority of the lightcone voxels. Additionally, although both our method and the neural ensemble method offer ``samples'' for uncertainty quantification, only our method can be interpreted as a proper Bayesian posterior.
Together with the sample-level power-spectrum agreement in Fig. \ref{fig:ps_kappa}, this behavior indicates that posterior-predictive intervals for non-linear summaries (e.g., peak counts, cross-correlations in COSMOS-Web) should be correctly calibrated on real data.

\section{Conclusion}

In this work, we introduce a first-of-its-kind 3D dark matter lightcone dataset tailored for WL mass mapping. Building on this resource, we propose a diffusion-based posterior sampler that reconstructs the 3D matter distribution from noisy 2D shear observations. Our approach significantly outperforms Neural Ensemble and Wiener filter baselines in 3D/2D correlation and power-spectrum fidelity on simulations designed to mimic JWST observations, and remains effective even under cosmology mismatch, indicating robustness to realistic modeling uncertainties. 
Important directions for future work include analyzing the sensitivity of reconstructions to the choice of simulation-derived prior, including different cosmological parameter settings and simulation codes.
By releasing our dataset, code, and evaluation protocol, we aim to create a reproducible testbed that opens 3D cosmological mass mapping to a far broader research community.
Our work brings humanity closer to seeing the dark universe in three dimensions, yielding reconstructions that not only reveal its structure with unmatched fidelity, but also unlock richer and more powerful downstream analyses.

\section{Acknowledgements}
The authors would like to thank Supranta Boruah, Bhuvnesh Jain, and Carolina Cuesta--Lazaro for helpful discussions. 
Part of this work was done at Jet Propulsion Laboratory, California Institute of Technology, under a contract with the National Aeronautics and Space Administration (80NM0018D0004) and funded through the President’s and Director’s Research and Development Fund (PDRDF). This work was also supported by the Stanback Innovation Fund and NSF Career award 2048237. 

{
    \small
    \bibliographystyle{ieeenat_fullname}
    \bibliography{main}
}

\input{sec/X_suppl}

\end{document}

%% file: sec/X_suppl.tex
\clearpage
\setcounter{page}{1}
\maketitlesupplementary

% \tableofcontents

\section{Visualization of Sample Quality}
\label{sec:sample_quality_visualization}

\begin{figure*}
    \includegraphics[width=\textwidth]{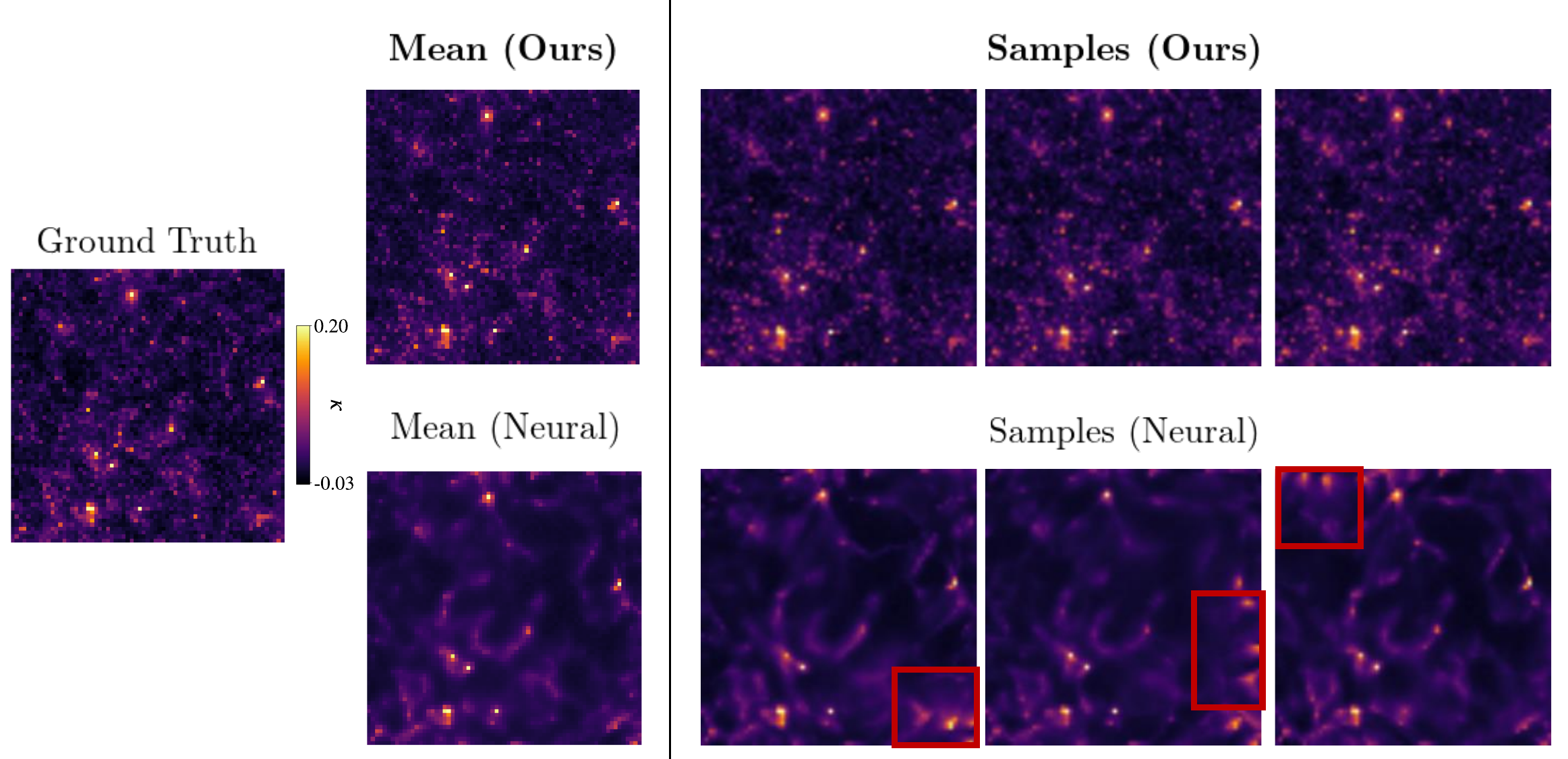}
    \caption{2D Convergence $\kappa$ Posterior Samples. Left: Ground-truth convergence field for a representative lightcone. Top row: Posterior mean and individual posterior samples from our diffusion-based method. Bottom row: Posterior mean and samples from the neural ensemble baseline. While both methods produce reasonable posterior means, the neural ensemble samples contain spurious small-scale artifacts that are washed out only when averaging. In contrast, our diffusion-based posterior produces visually coherent, artifact-free samples consistent with the statistics of the true $\kappa$ field.}
    
    \label{fig:k_samples}
\end{figure*}

\begin{figure*}
    \vspace{-.2in}
    \includegraphics[width=\textwidth]{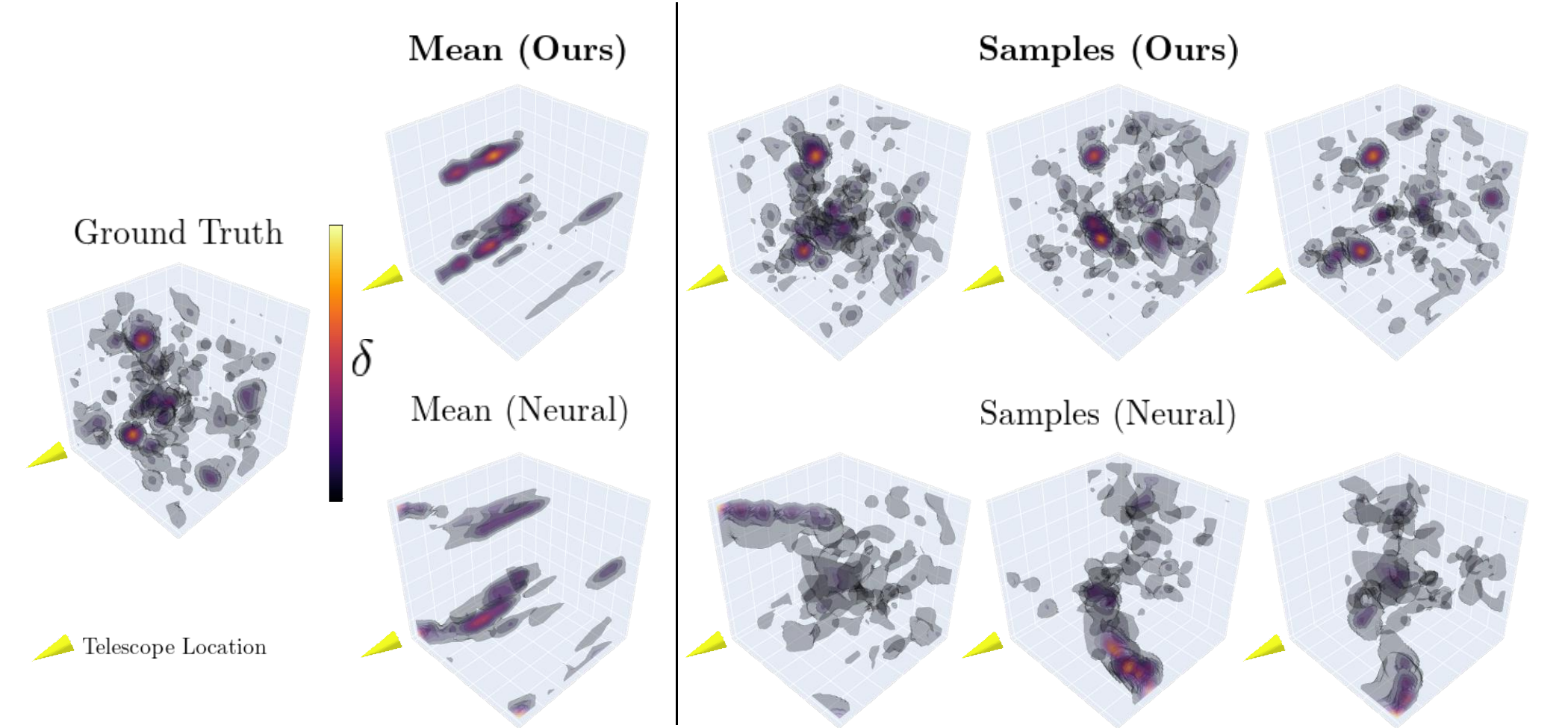}
    \caption{3D Overdensity $\delta$ Posterior Samples. Left: Ground-truth 3D overdensity for a representative volume. Top row: Posterior mean and individual samples from our diffusion-based sampler. Bottom row: Corresponding mean and samples from the neural ensemble. Neural-ensemble samples display artificial correlations along the line of sight, producing radially elongated structures not present in the true simulation. Our samples maintain realistic slice-to-slice variability and avoid spurious line-of-sight correlations; only the posterior mean exhibits expected radial blurring. These properties underscore the validity of our method’s 3D posterior samples for downstream non-Gaussian cosmological analyses.}
    
    \label{fig:od_samples}
\end{figure*}

A central goal of our diffusion-based posterior sampler is not only to produce a high-fidelity \emph{posterior mean} as a point estimate of the mass distribution, but also to generate \emph{realistic posterior samples} that reflect the variability and statistical structure of the underlying dark matter field. In this section, we provide qualitative visualizations of 2D and 3D samples (Figs. \ref{fig:k_samples} and \ref{fig:od_samples}) that complement the quantitative sample-quality diagnostics in the main text.

In the two-dimensional case, both our method and the neural ensemble produce posterior means that broadly recover the ground-truth convergence field $\kappa$. However, the individual samples reveal clear differences in sample quality. Samples from the neural ensemble contain spurious small-scale artifacts that are not present in the true $\kappa$ maps and that only disappear after averaging over many samples. In contrast, samples drawn from our diffusion-based posterior are relatively stable from realization to realization, without significant sample-wise artifacts. This indicates that our learned prior and sampling procedure preserve physically plausible structure at the level of individual convergence maps, not just in the mean.

The three-dimensional visualizations highlight an additional limitation of neural-ensemble-based uncertainty estimates. Neural ensemble samples exhibit spurious correlations along the line of sight: overdense regions are stretched into radially elongated features that persist across neighboring redshift slices, inconsistent with the statistics of the ground-truth overdensity field $\delta$ and with the radial power-spectrum behavior discussed in the main text. Our diffusion-based posterior samples, by contrast, maintain realistic slice-to-slice variability and remain relatively decorrelated across lens planes. The posterior mean naturally exhibits modest radial blurring, reflecting the intrinsic limits of 3D weak-lensing reconstruction, but individual samples themselves do not suffer from spurious line-of-sight coherence. These properties are crucial for downstream non-Gaussian analyses that operate directly on posterior samples, ensuring that ensemble-based statistics remain physically meaningful and well calibrated.

\section{Additional Recovery Results}

\begin{figure*}
    \includegraphics[width=\textwidth]{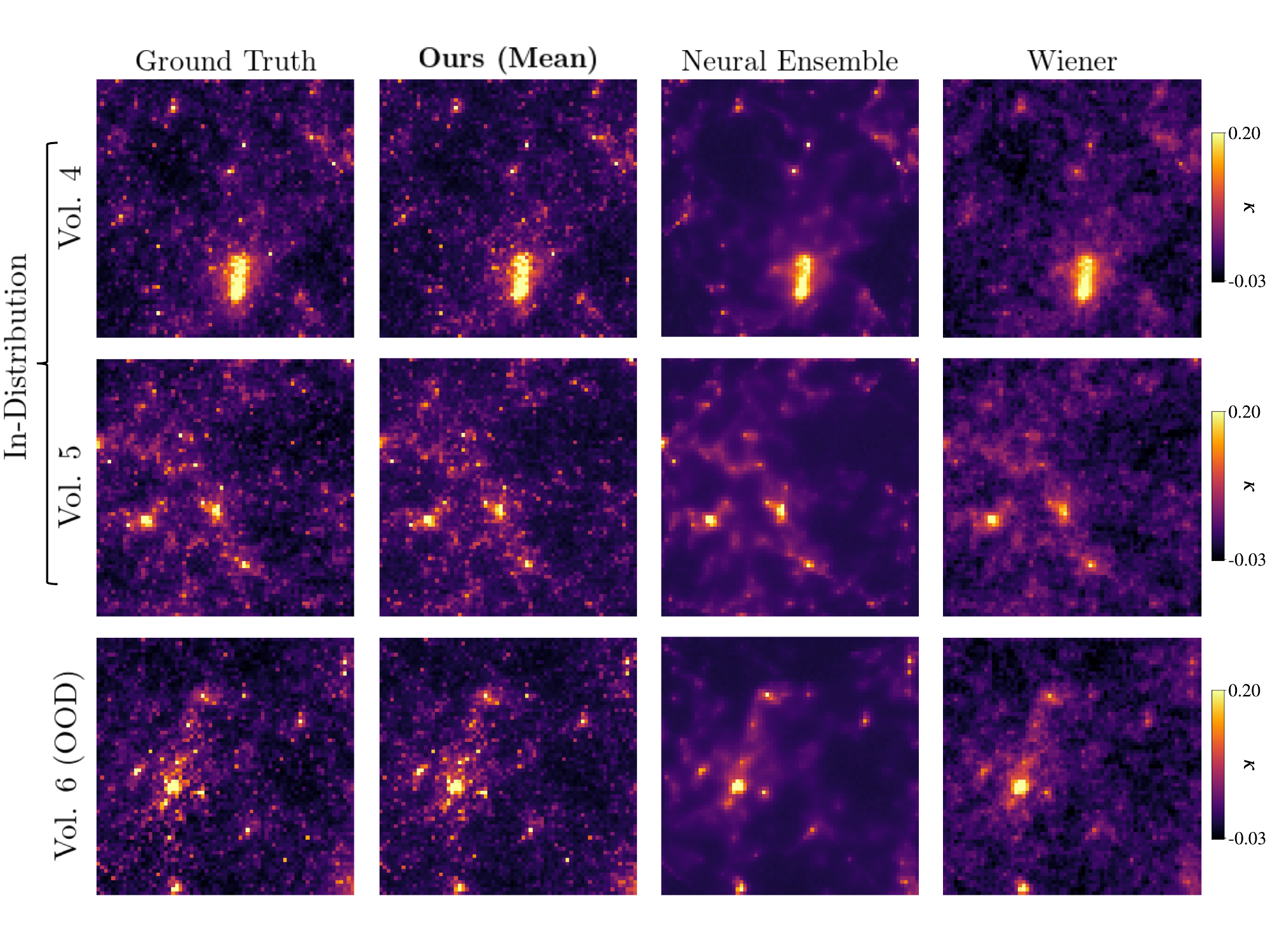}
    \caption{Additional 2D Recovery Results. For three additional simulated lightcones (Volumes 4, 5 and 6), we compare the reconstructed convergence $\kappa$ maps from our method, the Neural Ensemble baseline, and the Wiener filter against the ground truth. Consistent with the results in the main text, our method preserves high-contrast peaks and sharp filamentary features, whereas the baselines exhibit oversmoothing and reduced small-scale power. These trends persist across diverse structures and demonstrate the robustness of our 2D recovery performance.}
    
    \label{fig:recon2d_supp}
\end{figure*}

\begin{figure*}
    \vspace{-.2in}
    \includegraphics[width=\textwidth]{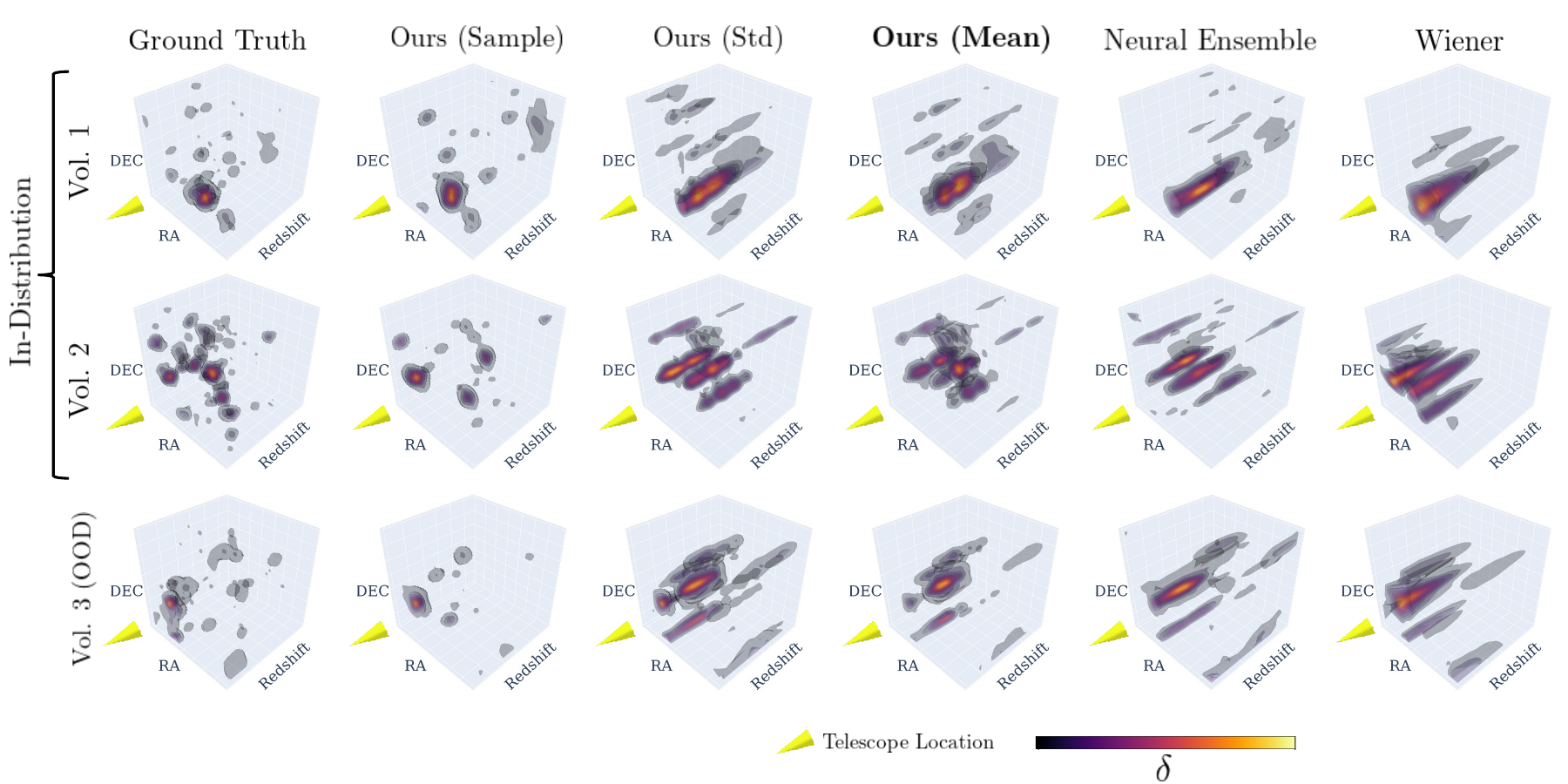}
    \caption{Additional 3D Recovery Results. We show full 3D overdensity $\delta$ reconstructions for Volumes 4, 5 and 6, comparing our diffusion-based posterior sampler with the Neural Ensemble and Wiener filter baselines. Our method more accurately localizes structures to the correct redshift slices and recovers finer-scale morphology along the line of sight. The baselines tend to smear overdense regions or miss small-scale features entirely. Together with Fig. \ref{fig:recon2d_supp}, these results confirm that the improvements observed in the main paper hold across all evaluated lightcones.}
    
    \label{fig:recon3d_supp}
\end{figure*}

To further validate the robustness of our method, we present recovery results for three additional simulated lightcones (Volumes 4, 5 and 6), expanding upon the qualitative and quantitative comparisons shown in the main paper. Similar to the reconstruction results in the main text, volumes 4 and 5 correspond to the in-distribution cosmology of the training set, while volume 6 corresponds to an out-of-distribution cosmology with massless neutrinos. Figure \ref{fig:recon2d_supp} and Figure \ref{fig:recon3d_supp} illustrate the 2D convergence $\kappa$ reconstruction and the full 3D overdensity $\delta$ recovery, respectively. Consistent with the three volumes shown in the main text, our diffusion-based posterior sampler reconstructs high-contrast cluster peaks and filamentary structure that closely match the ground truth, while the Neural Ensemble baseline tends to smooth overdensities and the Wiener filter over-regularizes fine features. These qualitative trends persist across all test volumes—including those with complex multi-scale structure—and indicate that our sampler reliably maintains small-scale power and accurate radial localization.

Table \ref{tab:recon_supp} provides quantitative cross-correlation metrics for these additional volumes. As before, we report (1) the blurred 3D correlation using $\sigma$ = 4 lensplanes, (2) the full-resolution 3D correlation, and (3) the projected 2D convergence correlation. Across all metrics, our method again achieves the highest agreement with the ground truth. 

\begin{table}
    \centering
    \begin{tabular}{ccccc}\toprule
         Vol.&  Method&  $\rho_\text{blur}^{3D} (\uparrow)$&  $\rho^{3D}(\uparrow)$& $\rho^{2D}(\uparrow)$ \\\midrule
         &  \textbf{Ours}&  \textbf{0.88}&  \textbf{0.25}& \textbf{0.93}\\
         4&   Neur. \cite{zhao2025revealing} &  0.85&  0.24& 0.92\\
         &  Wien. \cite{simon2009unfolding}&  0.84&  0.24& 0.90\\ \midrule
         &  \textbf{Ours}&  \textbf{0.85}&  \textbf{0.25}& \textbf{0.91}\\
         5&   Neur. \cite{zhao2025revealing} &  0.81&  0.18& 0.87\\
         &  Wien. \cite{simon2009unfolding}&  0.80&  0.22& 0.86\\ \midrule
         &  \textbf{Ours}&  \textbf{0.95}&  \textbf{0.24}& \textbf{0.88}\\
         6&   Neur. \cite{zhao2025revealing} &  0.91&  0.15& 0.82\\
         &  Wien. \cite{simon2009unfolding}&  0.86&  0.23& 0.77\\ \bottomrule
    \end{tabular}
    \caption{Additional quantitative comparison of recovery performance. We report 3D and 2D reconstruction quality for three additional simulated JWST-scale lightcone volumes: two in-distribution (Volumes 4 and 5) and one out-of-distribution with massless neutrinos (Volume 6). As in the main paper, we compute (1) the blurred 3D correlation $\rho^\text{3D}_\text{blur}$ after applying a Gaussian filter of width $\sigma = 4$ lensplanes along the line of sight, (2) the full-resolution 3D correlation $\rho^\text{3D}$, and (3) the 2D convergence correlation $\rho^\text{2D}$. Across all metrics and volumes, our diffusion-based posterior sampler achieves the highest agreement with the ground truth, outperforming the Neural Ensemble and Wiener filter baselines.}
    \label{tab:recon_supp}
\end{table}

\section{Detailed Resolution Analysis}

\begin{figure*}
    \includegraphics[width=\textwidth]{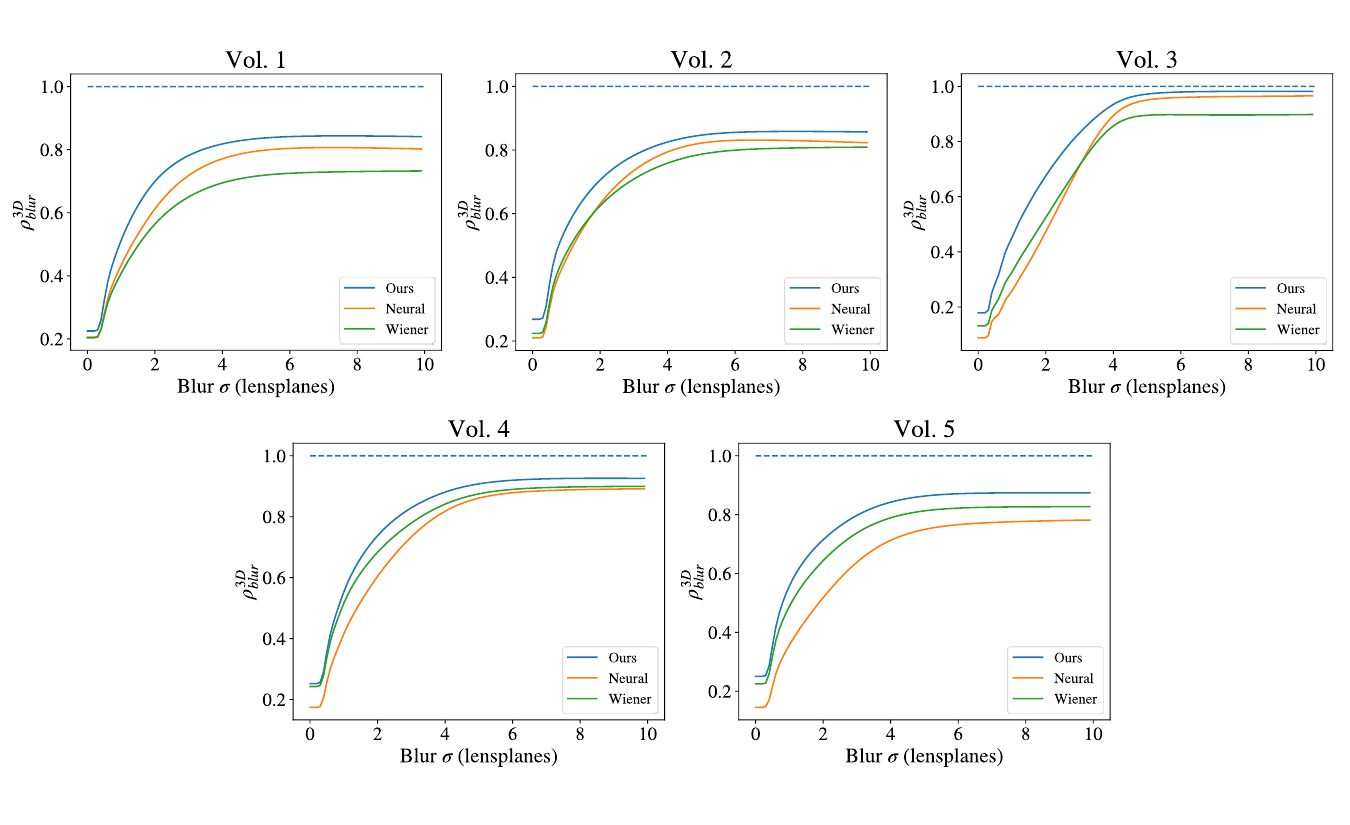}
    \caption{Resolution Analysis across radial blur scales. We evaluate the blurred 3D cross-correlation $\rho_\text{3D}^\text{blur}(\sigma)$ between each reconstruction and the ground truth as a function of the Gaussian blur width $\sigma$ (in lensplanes). Larger $\sigma$ corresponds to more aggressive smoothing along the line of sight. Across all five simulated volumes, our method achieves higher correlation than both baselines for every blur scale, demonstrating improved depth resolution at both fine and coarse radial smoothing regimes. The consistently superior performance indicates that our diffusion-based prior enhances radial structure recovery independently of the chosen blur width.}
    
    \label{fig:blur_res}
\end{figure*}

In the main paper, we evaluated 3D reconstruction fidelity using a single Gaussian blur scale ($\sigma = 4$ lensplanes) as a summary metric for radial resolution. Figure~\ref{fig:blur_res} shows the blurred 3D cross-correlation $\rho_\text{3D}^\text{blur}(\sigma)$ between each reconstruction and the ground-truth volume as a function of $\sigma$, where both volumes are convolved with the same Gaussian kernel and larger $\sigma$ corresponds to more aggressive smoothing along the line of sight. For all five test volumes, our method outperforms the baselines at every blur scale, demonstrating that the improvements provided by our diffusion prior are not tied to any particular choice of radial smoothing.

\section{Weak Gravitational Lensing Formalism}
\subsection{Deflection and Shape Distortion}
A more explicit derivation of \cref{eq:proj}–\cref{eq:forward_model} is presented below. Further details can be found in \cite{bartelmann2001weak, schneider2006weak}.

Consider a background galaxy (``source'') whose image is being weakly lensed by an extended 3D mass distribution (``lens''). The source is observed at angular position $\theta$ on the sky, whereas in the absence of lensing it would appear at $\boldsymbol{\beta}$. We therefore define the scaled deflection angle $\boldsymbol{\alpha}$ through the lens equation:  

\begin{equation}
 \label{eq:lens_eq}
 \boldsymbol{\beta} = \boldsymbol{\theta} - \boldsymbol{\alpha}.   
\end{equation}

For an extended 3D mass distribution, the deflection angle depends on the gravitational potential $\Phi$ encountered by the light ray along its path from source to observer. Under the Born approximation \cite{bartelmann2001weak}, this becomes the line-of-sight integral: 

\begin{equation}
    \label{eq:deflection}
    \boldsymbol{\alpha}(\boldsymbol{\theta}, w) = \frac{2}{c^2} \int_0^w dw' \frac{w - w'}{w} \nabla_\perp \Phi(w' \boldsymbol{\theta}, w').
\end{equation}

This expression is closely analogous to the Eikonal ray equation from optics, which describes the deflection of a ray by a refractive-index field \cite{atcheson2008time, born2013principles}. For convenience, we pull the transverse gradient $\nabla_\perp$ outside the integral and define the \textit{effective lensing potential} $\psi(\boldsymbol{\theta}, w)$, which represents the weighted integral of $\Phi$ along the line of sight: 

\begin{align}
    \psi(\boldsymbol{\theta}, w) &= \frac{2}{c^2} \int_0^w dw' \frac{w - w'}{w \cdot w'} \Phi(w' \boldsymbol{\theta}, w'), \\
    \boldsymbol{\alpha} &= \nabla_{\boldsymbol{\theta}} \psi
\end{align}

While gravitational lensing deflects the apparent position of light rays, our primary interest lies in how it distorts the ellipticity $e$ of a galaxy image (\cref{eq:ellipticity}). In the weak-lensing regime, a first-order expansion shows that the shear components $\gamma$ (\cref{eq:shapechange}) are given by the partial derivatives of the scaled deflection angle \cite{bartelmann2001weak}: 

\begin{align}
    \gamma_1 & \approx \frac{1}{2}(\boldsymbol{\alpha}_{11} - \boldsymbol{\alpha}_{22}) \\
    \gamma_2 & \approx \boldsymbol{\alpha}_{12} = \boldsymbol{\alpha}_{21}
\end{align}

Here, we adopt the notation $\boldsymbol{\alpha}_{ij} = {\partial\boldsymbol{\alpha}_i} / {\partial\boldsymbol{\theta}_j}$, and $ \boldsymbol{\alpha}_{12} = \boldsymbol{\alpha}_{21}$ follows from the fact that $\boldsymbol{\alpha}$ is a gradient field. Consequently, the shear can also be expressed in terms of the effective lensing potential: 

\begin{align}
    \gamma_1 &= \frac{1}{2}(\frac{\partial^2\psi}{\partial\theta_1^2} - \frac{\partial^2\psi}{\partial\theta_2^2}) \label{eq:g1} \\
    \gamma_2 &=  \frac{\partial^2\psi}{\partial\theta_1\partial\theta_2} \label{eq:g2}
\end{align}

Intuitively, instead of a point-like ray, one may picture a small ellipse representing a galaxy image. The gradient (first-order derivatives) of $\Phi$ determines the deflection $\alpha$ of the ellipse’s center, while the second-order derivatives of $\Phi$ describe the shear $\gamma$ that distorts its shape. To connect this potential to the underlying matter field, quantified by the density overdensity $\delta$, we now introduce the convergence $\kappa$:
\begin{align}
    \kappa &= \frac{1}{2} \Delta_{\boldsymbol{\theta}} \psi \label{eq:k1} \\
    &= \frac{1}{c^2} \Delta_{\boldsymbol{\theta}} \int_0^w dw' \frac{w - w'}{w \cdot w'} \Phi(w' \boldsymbol{\theta}, w') \\
    &= \frac{1}{c^2} \int_0^w dw' \frac{w'(w - w')}{w} \Delta \Phi(w' \boldsymbol{\theta}, w') \label{eq:k2}
\end{align}

In \cref{eq:k2} we assume that the $\Phi$ has negligible derivative in the radial direction. Next, applying the Poisson equation for gravitational potential in comoving coordinates, 

\begin{equation}
    \Delta \Phi = \frac{3H_0^2\Omega_m \delta}{2a}
\end{equation}

which leads directly to the weighted projection of the matter overdensity given in \cref{eq:proj}. To relate the convergence $\kappa$ to the shear $\gamma$, we first invert the Poisson equation given in \cref{eq:k1} using the method of Green's functions \cite{kaiser1993mapping, evans2022partial}: 

\begin{equation}
    \Tilde{\psi}(\boldsymbol{\theta}) = \frac{1}{\pi} \int_{\mathbb{R}^2} d^2\boldsymbol{\theta'} \ln |\boldsymbol{\theta} - \boldsymbol{\theta'}| \kappa(\boldsymbol{\theta'}) 
\end{equation}

Finally, plugging $\Tilde{\psi}$ into \cref{eq:g1} and \cref{eq:g2} yields \cref{eq:conv}.

\subsection{Photometric Redshift Errors}
When a survey relies on photometric redshifts, the projection defined in equation \cref{eq:proj} must be marginalized over the uncertainty in the source’s true redshift. For a galaxy with a redshift probability distribution 
$p(z)$, the expected convergence is therefore obtained by averaging  $\kappa(\boldsymbol{\theta},w)$ over this distribution

\begin{equation}
    \kappa(\boldsymbol{\theta}) = \mathbb{E}_{z \sim p(z), z = z(w)} \left[ \kappa(\boldsymbol{\theta}, w)  \right] 
\end{equation}
Conceptually, this replaces the upper limit in the line-of-sight projection with an expectation over all comoving distances consistent with the photometric redshift estimate. This leads to the standard expression in which the convergence is written as an integral over comoving distance
\begin{equation}
    \kappa(\boldsymbol{\theta}) = \frac{3H_0^2 \Omega_m}{2c^2} \int_0^w dw \frac{\delta(w \boldsymbol{\theta}, w)}{a(w)}g(w)w 
\end{equation}

Here, $g(w)$ denotes the lensing efficiency, which incorporates the probability that the galaxy lies behind a given lens plane. It is defined by

\begin{equation}
    g(w) = \int_w^{w_k} dw' \frac{w' - w}{w'}\left(p(z) \frac{dz}{dw} \right)_{z = z(w')}.
\end{equation}

Here $w_k$ is some reasonable upper limit on the source comoving distance. In practice, the efficiency function $g(w)$ can be precomputed for each galaxy at any set of lens-plane positions. To evaluate the required derivatives, such as 
$dz/dw$ and the mapping between redshift and comoving distance, we use the cosmology package \texttt{jax-cosmo}, which provides differentiable cosmological functions.

Throughout this work, we model the photometric-redshift distribution $p(z)$ as a Gaussian centered on the measured redshift,

\begin{equation}
    p(z) \sim \mathcal{N}(z, \sigma_z), \sigma_z = 0.11(1 + z)
\end{equation}
 
with the scatter $\sigma_{z}$ estimated from the JWST COSMOS-Web shear catalog \cite{scognamiglio2025highest, shuntov2025cosmos2025}. This Gaussian approximation is standard and has been adopted in several previous weak-lensing studies \cite{simon2009unfolding, leonard2014glimpse}.

\section{\texttt{Conicus3D} - Dataset Construction}

\texttt{Conicus3D} is a dataset of 3D lightcones, binned into lensplanes, derived from the \texttt{AbacusSummit} simulations \cite{garrison2021abacus, maksimova2021abacussummit}. We use the 25 \texttt{base} boxes with the AbacusSummit fiducial cosmology (\texttt{AbacusSummit\_base\_c000\_ph\{000-024\}}), as well as one \texttt{base} box from the cosmology corresponding to massless neutrinos (\texttt{AbacusSummit\_base\_c009\_ph000}). Each box contains two lightcones spanning approximately $28 \times 28$ deg$^2$, from which we crop smaller patches to train the diffusion model on a chosen survey footprint.

To construct the overdensity lensplanes, we follow the methodology in \cite{hadzhiyska2023synthetic} to compute the overdensity maps on spherical shells. Given a particle shell map with particle count $N_{ij}$ at pixel index $ij$, we compute the overdensity as: 

\begin{equation}
    \delta(ij) = \frac{\rho(ij)}{\Bar{\rho}} - 1
\end{equation}

where 

\begin{equation}
    \Bar{\rho} = (N_\text{part} / L_\text{box}^3) \Delta\Omega \chi_j^2 d{\chi_j}
\end{equation}

is the mean density. Here $N_\text{part} = 6912^3$ is the number of particles in a simulation box, $L_\text{box} = 2$ Gpc/$h$ is the simulation box length, $\Delta \Omega$ is the solid-angle pixel area, and $\chi_j$ and $d{\chi_j}$ are the shell center and width, respectively.  The density per pixel is therefore 

\begin{equation}
    \rho(ij) = \frac{N_{ij}}{dV_j} = \frac{N_{ij}}{\Delta\Omega\chi_j^2 d\chi_j}.
\end{equation}

After computing an overdensity map on each shell, we combine the shells into 20 lensplanes using a weighted average. These planes span comoving distances from approximately $290$ to $3580$ ($z \approx 2$) Mpc/$h$. Each lensplane is assigned an effective scale factor and comoving distance; values for a subset of these planes are provided in Table~\ref{tab:lensplanes}.

\begin{table*}
    \vspace{1in}
    \centering
    \begin{tabular}{ccccc}\toprule
         Plane Index&  Redshift $z$&  Scale Factor $a$&  Center (Mpc/$h$)& Width (Mpc/$h$)\\\midrule
         0&  0.13&  0.88&  372.25& 164.5\\
         2&  0.25&  0.80&  701.25& 164.5\\
         6&  0.52&  0.66&  1359.25& 164.5\\
         12&  1.04&  0.49&  2346.25& 164.5\\
         19&  1.92&  0.34&  3497.75& 164.5\\ \bottomrule
    \end{tabular}
    \caption{Geometry of selected \texttt{Conicus3D} lensplanes. For reference, we list the redshift, scale factor, comoving distance, and physical width for several representative lensplanes used in the \texttt{Conicus3D} dataset. These planes are derived from overdensity shells computed following the methodology of \cite{hadzhiyska2023synthetic} and subsequently binned into 20 evenly spaced lensplanes in comoving distance. The planes span the full radial range of the lightcone (approximately 290–3580 Mpc/$h$, corresponding to $z \approx 2$), and their geometry is used both in constructing the dataset and in defining the line-of-sight distances relevant for weak-lensing projections. This subset illustrates the non-uniform evolution of redshift and scale factor across the lightcone and provides context for the depth-dependent analysis presented in the supplement.}
    \label{tab:lensplanes}
\end{table*}

\begin{algorithm*}[t]
\caption{Modified Decoupled Annealing Posterior Sampling}
\label{alg:daps}
\begin{algorithmic}[1]
\Require Observation $\gamma$, diffusion prior $p(\delta)$,
         weak-lensing forward model, annealing schedule
         $\{\sigma_T, \dots, \sigma_0\}$, power spectrum $P(k)$
\State Initialize $\delta_T \sim \mathcal{N}(0, \sigma_T^2 \mathbf{I})$
\For{$t = T, T-1, \dots, 0$}
    \State \textbf{Prior projection via diffusion model:}
    \State $\hat{\delta}_0 \gets \textsc{ReverseDiffusion}(\delta_t, \sigma_t)$
    \State \textbf{Likelihood-guided refinement (Langevin dynamics):}
    \State Define prior covariance $\boldsymbol{\Sigma}$ using Fourier diagonal $P(k)$
    \State Sample $\delta_{0 \mid \gamma}$ with Langevin updates targeting
           $\log p(\gamma \mid \delta) + \log \mathcal{N}(\delta; \hat{\delta}_0, \boldsymbol{\Sigma})$
    \If{$t > 0$}
        \State Draw $\epsilon \sim \mathcal{N}(0, \mathbf{I})$
        \State $\delta_{t-1} \gets \delta_{0 \mid \gamma} + \sigma_{t-1} \, \epsilon$
    \EndIf
\EndFor
\State \Return samples $\{\delta_0\}$ approximating $p(\delta \mid \gamma)$
\end{algorithmic}
\end{algorithm*}

\section{Decoupled Annealing Posterior Sampling}
\label{sec:daps_supp}

To draw samples from the posterior distribution $p(\delta \mid \gamma)$, we use a modified version of Decoupled Annealing Posterior Sampling (DAPS) \cite{zhang2025improving}. DAPS constructs a sequence of noise-annealed posteriors $p(\delta_t \mid \gamma)$ indexed by a decreasing noise scale $\sigma_t$. Sampling proceeds from a high-noise initial distribution toward the target posterior, alternating between (1) projecting the current sample toward the learned prior using the diffusion model, and (2) refining it using likelihood guidance from the weak-lensing forward model.

At each stage, the diffusion model provides a clean estimate $\hat{\delta}_0$ consistent with the data-driven prior. This estimate is then refined using Langevin dynamics that incorporates both the WL likelihood and a Gaussian prior term. Unlike the standard DAPS formulation, we replace the diagonal covariance of this Gaussian prior with a Fourier-space covariance derived from the estimated matter power spectrum. This modification enforces statistical isotropy and translation invariance and helps ensure that radial and angular correlations of the overdensity field are preserved throughout sampling.

After refinement, the sample is re-noised according to the next noise level $\sigma_{t-1}$, enabling a coherent annealing schedule. Iterating this process yields approximate samples from $p(\delta \mid \gamma)$, which accurately capture both small-scale structure and the full posterior variability needed for downstream cosmological analyses.

\section{Broader Applicability of the Spectral Covariance Formulation}
\label{sec:spectral_applicability}

The prior covariance in our formulation (Eq.~7) is diagonalized in Fourier
space by the power spectrum, a structure that arises naturally whenever the
field of interest is approximately translation-invariant. This property is
common across scientific inverse problems.

In adaptive optics, the Kolmogorov power spectrum characterizes atmospheric
wavefront phase covariance, and phase screens are standardly generated by
spectral filtering in Fourier space~\cite{lane1992simulation}. In
geophysical inversion, subsurface properties such as seismic velocity and
porosity are modeled as stationary fields with spectrally defined
covariance for posterior sampling in tomography and reservoir
characterization~\cite{hansen2006linear}. In CMB analysis, the angular
power spectrum $C_\ell$ defines a harmonic-diagonal signal covariance for
Wiener filtering and constrained realizations of the microwave
sky~\cite{elsner2013efficient}.

In each case, translation invariance permits efficient spectral
representation of the prior. We believe our spectral sampling strategy could
be adapted to improve diffusion-based posterior sampling in these and similar
inverse problems, as properly modeling the covariance structure allows the
sampler to more faithfully represent the learned prior distribution during
inference.